# Magnetoresistance of vertical Co-graphene-NiFe junctions controlled by charge transfer and proximity-induced spin splitting in graphene

P. U. Asshoff[1], J. L. Sambricio[1], A.P. Rooney[2], S. Slizovskiy[3], A. Mishchenko[1], A.M. Rakowski[2], E.W. Hill[3], A. K. Geim[1], S. J. Haigh[2], V. I. Fal'ko[1,3], I. J. Vera-Marun[1], I. V. Grigorieva[1*]

[1]*School of Physics and Astronomy, University of Manchester, Oxford Road, Manchester M13 9PL, UK*

[2]*School of Materials, University of Manchester, Oxford Road, Manchester M13 9PL, UK*

[3]*National Graphene Institute, University of Manchester, Manchester M13 9PL, UK*

**Graphene is hailed as an ideal material for spintronics due to weak intrinsic spin-orbit interaction that facilitates lateral spin transport and tunability of its electronic properties [1-3], including a possibility to induce magnetism in graphene [4-9]. Another promising application of graphene is related to its use as a spacer separating ferromagnetic metals (FMs) in vertical magnetoresistive devices [10-20], the most prominent class of spintronic devices widely used as magnetic sensors. In particular, few-layer graphene was predicted [10-12] to act as a perfect spin filter. Here we show that the role of graphene in such devices (at least in the absence of epitaxial alignment between graphene and the FMs) is different and determined by proximity-induced spin splitting and charge transfer with adjacent ferromagnetic metals, making graphene a weak FM electrode rather than a spin filter. To this end, we report observations of magnetoresistance (MR) in vertical Co-graphene-NiFe junctions with 1 to 4 graphene layers separating the ferromagnets, and demonstrate that the dependence of the MR sign on the number of layers and its inversion at relatively small bias voltages is consistent with spin transport between weakly doped and differently spin-polarized layers of graphene. The proposed interpretation is supported by the observation of an MR sign reversal in biased Co-graphene-hBN-NiFe devices and by comprehensive structural characterization. Our results suggest a new architecture for vertical devices with electrically controlled MR.**

Following the successful development of graphene-based lateral spintronic structures [1-8], the implementation of graphene as a spacer in vertical magnetic tunnel junctions (MTJ) has become a subject of intense interest [13-20]. Up to now, theoretical proposals [10-12] for graphene's role in MTJs focused on the so-called 'K-point spin filtering' expected in ideally lattice-matched single-crystalline ferromagnet–graphene–ferromagnet (FM-G-FM) structures and attributed to matching spin-polarized bands in the ferromagnet and the electronic states in the graphene treated as a tunnelling barrier. This mechanism was also used to interpret the MR sign inversion observed in conventional Ni-Al$_2$O$_3$-Co [13] and Ni-MgO-Co [14] tunnel junctions where the Ni electrode was passivated by CVD grown (epitaxial) mono- [13] or few-layer [14] graphene. However, despite several attempts, no significant spin filtering could so far be achieved in vertical magnetoresistive devices where graphene was used as a spacer between two FM electrodes without an additional insulating barrier [15-20], in contrast to theory predictions [10-12]. The most likely reason for this is that the reported FM-G-FM sandwiches (especially those fabricated by a combination of graphene transfer and deposition of FM films) were in fact van der Waals heterostructures [21,22] rather than truly lattice-matched crystals. In this case, as we show below, graphene plays a different role, itself becoming a source of spin-polarized electrons due to the interplay of (i) its doping by charge transfer from the FM metal and (ii) proximity-induced spin splitting in graphene.

To explore such a possibility, one needs to create contamination- and oxidation-free G-FM interfaces, but without the requirement of lattice matching between G and the FMs. To this end, we have fabricated vertical FM-G-FM' devices where ultimately clean G-FM interfaces were obtained by depositing the ferromagnetic metals on the two sides of a suspended graphene membrane, thereby preventing oxidation, minimizing the number of fabrication steps and limiting the exposure of the devices to solvents during preparation (here FM=Co and FM'=$Ni_{0.8}Fe_{0.2}$ alloy). Details of sample fabrication and characterization are given in Methods and Supplementary Figures S1-S3,S8. Briefly, an exfoliated graphene (or hBN) flake was suspended over a 3-4 µm diameter circular aperture in a 100 nm-thick $SiN_x$ membrane using a dry transfer method [23]. The FM electrodes were then evaporated onto the suspended flake from the top and bottom side as shown schematically in Fig. 1(a) and Supplementary Fig. S1 (a similar approach was used in ref. [15], but our method is different in several important aspects, as described in Methods and Supplementary Note 1). After the first (top) deposition, the metal at the interface was protected from oxidation by graphene, as the latter is known to act as an impermeable barrier for all gases and liquids [13,24], which also implies that pinholes are virtually non-existent.

To test our fabrication procedure, we first made a device where Co and $Ni_{0.8}Fe_{0.2}$ (Py) electrodes were separated by bilayer hexagonal boron nitride (hBN) instead of graphene. Atomically thin hBN has been shown to act as a high-quality insulating barrier in vertical transistors [25], lateral spin valves [26] and, most recently, in magnetic tunnel junctions (MTJs) [27]. Figures 1b,c show magneto-transport characteristics of our FM-hBN-FM' device. These are consistent with classical MTJ behavior: The two electrodes are magnetically separated and switch independently, with a TMR value of ≈ 1% at a temperature $T$ = 10 K. Using Jullière's formula [28] $TMR = 2P_1P_2/(1 - P_1P_2)$ and a simplifying assumption $P_1 = P_2 = P$ (equal spin polarization for both FM electrodes) we obtain an average spin polarization of the ferromagnets $P$~7%. These characteristics compare very favorably with $P$ ~ 0.05 - 0.25% and TMR ~0.3-0.5% reported for Co-hBN-NiFe MTJs based on hBN grown by chemical vapor deposition (CVD) and transferred onto the FM [29] and are of the same order as the more recent work [27] where the hBN barrier in Fe/hBN/Co MTJs was grown directly on Fe. The temperature-dependent MR shown in the inset of Fig. 1(b) is also in agreement with the usual TMR behavior [28-31]: the MR decreases by 40% as the temperature is increased from 5 K to 300 K, which is commonly attributed to magnon excitations [30,31]. Furthermore, the non-linear $I(V)$ characteristics (Fig. 1c) and the resistance × area (RA) product of 6.2 kΩ·µm² at 10 K agree with previous measurements on a 2-layer hBN tunnel barrier [32] and the corresponding d$I$/d$V$ (inset in Fig. 1c) is in agreement with earlier studies of Co-based MTJs with AlO$_x$ barriers [33]. Finally, we have checked that our fabrication method does not lead to unintentional contacts between the top and bottom FM films by preparing several test samples without the graphene-covered aperture and found resistances >10 MΩ, confirming that the current flows through the junction area only.

Having proven the technology, we fabricated a set of ten Co-G-Py structures, where FM electrodes are separated by mono-, bi-, tri- or four-layer graphene, and studied their vertical magnetoresistance. The sample design is shown in Fig. 1(a). Typical variations of the MR as a function of in-plane magnetic field, $B$, at different temperatures, $T$, are shown in the top row of Fig. 2a and Supplementary Fig. S5. Similar to the hBN-based MTJ device, all graphene-based junctions showed positive MR, with plateau regions in the $R(B)$ traces, as expected in a regular MTJ device, indicating that already a monolayer of graphene is sufficient to decouple the switching of the two FMs. This is consistent with earlier reports, where a similar decoupling of the ferromagnetic electrodes has been observed for a monolayer graphene spacer [15,16]. At the same time, the resistance of all devices was quite low (between 2 and 20 Ω) and $I(V)$ characteristics were linear (Supplementary Fig. S4), corresponding to metallic behavior. This is in stark contrast to some previous reports [15-17] where graphene-based structures showed tunneling behavior; e.g., ref. [17] reported junction RA products in excess of 35 kΩ·µm$^2$, probably due to partial oxidation or residual contamination of the ferromagnetic films during the transfer of graphene in ambient conditions. Concerning the origin of MR in our devices, we used angle-dependent measurements to verify that it was not due to anisotropic magnetoresistance (see Supplementary Note 3 for details).

In terms of RA and MR values, there was no correlation between the number of graphene layers, $N$, separating the FM electrodes and the device behavior (top row of Fig. 2a, Fig. 3a and Supplementary Fig. S5), despite the identical fabrication conditions. For example, for three $N$=2 devices the measured RA products were 17, 29 and 309 Ω·µm$^2$, with MR = 1.03%, 0.66% and 0.10%; for two $N$=4 devices these were 38 and 136 Ω·µm$^2$, or 0.42% and 0.12%, respectively. A similarly large spread of characteristics was found for $N$ =3 (RA = 23, 170, 193 Ω·µm$^2$ with MR = 0.9%, 0.30%, 0.12%) and $N$ =1 (RA = 110, 175 Ω·µm$^2$ with MR = 0.09%, 0.2%) junctions. The contribution from the resistance of gold leads, ~1Ω, could only play a role for two devices with the smallest $R$ (see above). Furthermore, the large average value of RA across all samples, 120 Ω·µm$^2$ (much higher than for a typical metal) and the almost temperature-independent RA for most devices, including an increasing RA at low $T$ for $N$=3 and 4 in Fig. 3a, are strong indications that the measured $R$ of the devices is determined by the resistance of G-FM interfaces and not the metallic leads [34].

The surprisingly wide spread of (magneto)transport characteristics of our devices and the apparent lack of dependence on $N$ can be understood if we recall that the deposition of relatively thick, ~150 nm, metallic films on suspended, atomically thin graphene membranes is likely to result in significant strains in the devices, evident from their topography at different stages of preparation, e.g., the 'Mexican hat' shape shown in Supplementary Fig. S2. Strains will inevitably affect, locally, the atomic-scale separation between the FM and the adjacent graphene layer. To relieve such strains and further improve the interface quality, we annealed several representative devices with different $N$ at 300 °C in Ar/H$_2$ atmosphere for up to 10 h. Annealing under such conditions is known to significantly improve the interfaces in van der Waals heterostructures [23,35], and for our devices was confirmed by cross-sectional transmission electron microscopy (see below). This resulted in a strong change in both RA and MR, leading to a consistent behavior of all devices – c.f. Figs. 3a and 3b. The RA of all samples dropped significantly after annealing, from 17 – 196 Ω·µm$^2$ to 8 – 51 Ω·µm$^2$ at 10 K, and its temperature dependence became metal-like for all devices, unlike the almost flat or slightly increasing RA at low $T$ for the as-prepared samples. The change in magnetoresistance was even more dramatic: For devices with $N$ =2 and 3 MR changed sign and in some cases increased in absolute value (e.g., for $N$ =3 at $T$ =10 K from +0.30% before annealing to -0.6% after), see Fig. 2a,b. For $N$ =4 it dropped from MR = 0.4% to almost zero (Supplementary Fig. S6). Only for the monolayer device, $N$ =1, MR remained small and positive, slightly increasing from 0.1% to 0.14%.

Another change in device behavior becomes clear from comparison of the coercivities of the FM layers before and after annealing (Fig. 2a): these increased significantly, especially for $N$=2 and $N$=3, indicating more uniform magnetization in our polycrystalline FM films. It is therefore natural to attribute the increase in coercivity to relaxation of stresses both in the FMs and at the interface between a FM and the adjacent graphene layer. We note that the annealing temperature was too low to expect any significant recrystallization [36] and, indeed, no noticeable changes in grain sizes could be seen when the FM films were observed through transparent graphene in a scanning electron microscope (SEM), see Supplementary Fig. S3. Nevertheless, SEM observations of test samples (only one FM deposited on graphene, before and after annealing) did show signs of rearrangements of the individual crystallites in our polycrystalline FM films after annealing (Supplementary Fig. S3), which can be attributed to strain relaxation [36]. Relaxation of strains was also evident from atomic force microscopy (AFM) imaging of the junctions' topography after annealing (typical example is shown in Supplementary Figure S2): The 'Mexican hat' shape is no longer present (unsurprisingly, some sagging of the suspended area remains).

To confirm that the implied structural changes at the graphene-FM interfaces indeed take place in our Co-G-Py junctions, two representative devices before and after annealing were examined using cross-sectional transmission electron microscopy (TEM) – see Supplementary Fig. S8 and the associated discussion. This revealed the presence of significant, several nm-wide, voids between graphene and the ferromagnetic films before annealing, consistent with the observed large variations in resistance (a few voids are indicated by arrows in Supplementary Fig. S8). The voids effectively determine the MR before annealing, leading to a situation where the interface acts more like a 'bad' tunnel barrier (even though the details of the interface are device specific in this case). After annealing the voids disappeared, resulting in much smoother interfaces and uniform contact between graphene and the FMs, again consistent with the changes in junction resistance. Importantly, the annealing did not affect the chemical composition of the devices, that is, all constituent layers (Au, Ti, Co, Py) remained well defined, with no signs of interdiffusion (Supplementary Fig. 8).

Just an improved contact between graphene and FMs cannot account for the main feature in the behavior of our devices: the sign change of MR after annealing. The widely discussed 'K-point spin filtering', suggested by theory for both graphene-FM and hBN-FM epitaxial interfaces [10-12] and used to interpret a number of experimental findings [13,14,16,20], cannot explain the sign change of MR in our case and, more importantly, is not applicable to our devices. Firstly, the polycrystalline nature of our FM films and the corresponding lack of epitaxy with graphene (or hBN) are contrary to the theory assumptions. Secondly, the RA product in our experiments is 5 orders of magnitude greater than the theoretical value, < 0.003 Ω μm$^2$, obtained for 'K-point spin filtering' and attributed to hybridization and covalent bonding [11]. Therefore, we draw attention to two features specific to the FM/graphene interface which have not been considered previously in the context of MTJs: doping due to contact with a metal [37,38] and exchange splitting due to proximity to a ferromagnet that gives rise to a difference in the density of states (DoS) between spin-up and spin-down carriers in graphene [5-9]. Triggered by annealing, the conformation and improved contact between a FM electrode and the adjacent graphene layer results in effective decoupling of graphene layers within a van der Waals-bonded few-layer crystal, as was seen e.g. for twisted graphene bilayers under annealing [39, 40], producing two electronically decoupled thinner 'graphenes'. Subsequently, these two separately doped and exchange proximitised graphene layers become spin-polarised electrodes defining the device behavior, with a relatively small value of MR but qualitatively new features.

Here, we emphasize an important difference between the studied van der Waals structures and epitaxial G-FM or FM-G-FM' heterostructures. In vdW structures, the absence of lattice matching between the FMs and the adjacent graphene layers prohibits hybridization of the electronic bands of graphene and Co/Py (hybridization would effectively extinguish the Dirac spectrum of graphene). There is consensus between theoretical studies [10,12,38] that whether or not hybridization occurs depends strongly on two related factors, lattice matching and the distance, $d_{FM/G}$, between the metal and carbon atoms. Hybridization is likely to occur for $d_{FM/G}$ ~2 Å in lattice-matched structures, such as epitaxial G-Co films [41], or epitaxial G-Ir intercalated by Co [42]). For example, a recent STM and ARPES study of graphene in direct contact with Ni [43] emphasized that precise lattice matching is needed for a shorter $d_{FM/G}$ and hybridization. In contrast, for typical vdW spacings ($d_{FM/G} \geq 3$ Å) graphene's linear spectrum is preserved and proximity-induced spin splitting is expected in the same way as when graphene is in contact with a ferromagnetic insulator [5-9]. In our devices, we were able to measure the graphene/metal distance, $d_{FM/G}$, directly using cross-sectional TEM (Supplementary Fig. S8). This showed that for the Co-graphene interface (after annealing) $d_{FM/G} = 0.39 \pm 0.06$ nm and for Py-graphene $d_{FM/G} = 0.34$ nm $\pm 0.09$ nm, see Supplementary Note 4. The absence of hybridization also prevents direct RKKY coupling of the ferromagnetic electrodes via graphene considered in Refs. [41,42]. In our vdW structures, the distance between two ferromagnetic layers, ~7.4Å is almost twice longer than in epitaxial structures [41,42], therefore, the interlayer RKKY interaction in our devices can be neglected, in agreement with the observed independent switching of the polarization of the two ferromagnetic films in our Co-G-Py devices.

In this situation, unique to graphene, the sign of the spin polarization (majority/minority carriers) should depend on the sign and level of doping (even for the same exchange splitting). This is because for spin-split bands in gapless graphene the type (p- or n-) of doping would determine which spin components (up- or down-) would have larger DoS at the Fermi level, as illustrated schematically in Fig. 2c. To evaluate the doping in our devices, we prepared test samples as described in Supplementary Note 5 and measured gate-dependent charge transport in the vicinity of Co and Py contacts. This showed that graphene in contact with Co is n-doped and graphene in contact with Py is p-doped, with corresponding shifts of the Fermi energy, $E_F$, from the charge neutrality point ≈ -100 meV and +190 meV, respectively (Supplementary Note 5). A scenario corresponding to such doping is illustrated in Fig. 2c for the charge transfer and proximitised exchange splitting generated by the spin-dependent DoS of Co, Py (drawn schematically with partly occupied 3d and 4s states) and the few-layer graphene spacer. In this simple picture, at $E_F$ spin-up electrons represent the majority in n-doped graphene and spin-down electrons are the majority in p-doped graphene. It follows directly that the sign of MR will be negative.

Note that the proposed interpretation also explains why, upon annealing, MR increases (slightly) for $N=1$ but decreases and changes sign for $N \geq 2$. A monolayer graphene can conform only to one of the FM layers and the small increase in MR can be explained by an improved contact on one side, while few-layer graphene crystals 'split' and electronically decouple, as described above, and become differently polarized, resulting in a negative MR. The sign change for devices with $N = 2$ and 3 is very clear and occurred in all studied devices. As for the devices with N=4, they showed a drop in resistance after annealing, consistent with the behavior for N=1, 2 and 3, but the sign reversal of MR was not observed (Supplementary Fig. S6). There are at least two likely reasons for the observed difference between devices with $N = 2,3$ and $N=4$, as explained in detail in Supplementary Note 3. Briefly, N=4 graphene is stiffer and one can expect a less effective conformation at the interfaces with the FM films [45]. Also, in the likely event that a tetralayer 'separates' into two bilayers, a smaller metal-induced shift of the Fermi level, $E_F$, compared to the bi- or trilayer devices should result in smaller spin polarizations, very small MR and the absence of (or an incomplete) MR sign reversal, in agreement with our observations.

To further verify the proposed model, we performed additional experiments on Co-G-NiFe and Co-G-hBN-NiFe devices using bias spectroscopy. First, we investigated the bias dependence of MR in Co-G-NiFe devices (with $N =2$ and 3) both before and after annealing (Fig. 4). In contrast to standard MTJs, the role of increasing bias, $V_b$, in our devices is not only to shift the chemical potentials on the two sides of the barrier but also to change graphene doping (as shown schematically in Fig. 4), like in a graphene-based vertical transistor [25]. Due to low carrier density in graphene, it is possible to change its doping substantially using bias voltage [25,46], and even reverse its sign (e.g. from n-doping to p-doping), leading to a change of sign of MR. This is exactly what we observe in our devices, exemplified by the measurements shown in Fig. 4 for a Co-G-Py junction with a trilayer graphene spacer. We observed a change of MR sign from negative to positive for $V_b > +120$ mV (electrons flow from Py to Co), and a notable asymmetry in the rate of MR decrease for positive and negative $V_b$. For negative $V_b$ we only observe a monotonic decrease of MR with increasing bias, as is common in standard MTJs [31,49]. For positive $V_b$ this usual behavior is outweighed by the changing spin polarization and the reversal of MR sign. Note that, in contrast, a change of the Fermi level in the Co electrode itself which could lead to an MR sign reversal, would require $V_b \sim 800$ meV, as shown for Co/SrTiO$_3$/La$_{0.7}$Sr$_{0.3}$MnO$_3$ MTJs [47]. In our case we were able to tune the MR using much smaller bias voltages: from MR = - 0.6% at $V_b = 0$ to + 0.2% at $V_b = +230$ mV, with the sign change at $V_b \approx +120$ mV.

The bias dependence of Co-G-hBN-Py junctions (Fig. 5 and Supplementary Fig. S7) is even more straightforward to understand and is in full agreement with MR($V_b$) for our few-layer graphene devices. The MR at zero bias is positive and remains positive after annealing. This implies that the spin polarization in Co-proximitized, n-doped graphene is the same as at the Co-hBN interface [49] (cf. the positive MR in Fig. 1 for the Co-hBN-Py device). As the bias is increased to positive values (driving spin-polarized electrons from Py to Co) the Fermi level in graphene is first shifted towards the neutrality point and then into the valence band. Accordingly, the MR decreases and becomes negative (changes sign), on average, at $V_b \approx 80$ mV (for the device shown in Fig. 5 the sign change occurs at $V_b \approx +60$ mV while for the other two Co-G-hBN-Py devices, see Supplementary Fig. S7, the MR changes sign at $V_b \approx +80$ and $+105$ mV). We note that a qualitatively similar result (but negative MR) was recently reported for NiFe-G-hBN-Co MTJs [48], although the reason for the suggested negative spin polarization at the NiFe-graphene interface was not identified. An explanation follows directly from our experiments as being due to p-doping of graphene by NiFe.

The bias values corresponding to the change of MR sign both in Fig. 4 and 5 are remarkably similar to the doping-induced shift of $E_F$ in the graphene layer adjacent to Co, $\approx 100$ meV, as measured in our separate experiment (Supplementary Note 5). At first sight this is surprising, as one can expect a certain electrostatic 'back-screening' due to the presence of the second metal layer (Py) close by. To clarify whether such 'back-screening' indeed plays a role, we performed detailed analysis of charge transfer in metal-graphene multilayers (Supplementary Note 6). This showed that, while the shifts of $E_F$ in graphene layers adjacent to Co and Py were smaller than measured in our gate-dependent charge transport experiments (Supplementary Note 5), for the parameters corresponding to our devices, the bias-induced change of doping polarity in graphene adjacent to Co (from n- to p-doping) occurred at $V_b \sim +50$ to $+100$ mV, in excellent agreement with experiment.

Note that the larger value of MR for Co-G-hBN-Py devices compared to Co-G($N$=1)-Py (Fig. 2a,b) can be explained by two factors. Firstly, a significant contribution in this case comes from spin polarization at the hBN/NiFe interface, which from our experiments on Co-hBN-Py devices is estimated as ~7%. Secondly, and likely more importantly, the doping level of monolayer graphene in this case is well defined (n-doping) and an appreciable spin polarization can be expected. In contrast, in Co-(N=1)G-Py devices, doping of graphene in likely to be weak and poorly defined as graphene is in proximity to both Co and Py that dope in opposite sense. This explains that the maximum MR in the latter case is just 0.14%.

Our simple model allows us to estimate the value of the exchange splitting, $E_{ex}$, in graphene under the simplifying assumption that $E_{ex}$ is the same for Co- and NiFe-proximitised graphene. The MR and the spin polarizations are related through Jullière's formula [28] $MR = 2P_{Co}P_{Py}/(1 - P_{Co}P_{Py})$, where $P_{Co/Py}$ is the spin polarization in Co- or NiFe-proximitised graphene. The spin polarization in graphene, $P$, is related to $E_{ex}$ by

$$|P| = \frac{|D_\uparrow(E_{F,\uparrow}) - D_\downarrow(E_{F,\downarrow})|}{D_\uparrow(E_{F,\uparrow}) + D_\downarrow(E_{F,\downarrow})} = \frac{E_{ex}}{2|E_F|}$$

Here the density of states for spin-up and spin-down electrons are $D_{\uparrow/\downarrow}(E_{F,\uparrow/\downarrow}) \propto E_{F,\uparrow/\downarrow}$ [44] and the corresponding Fermi energies $E_{F,\uparrow/\downarrow} = E_F \pm E_{ex}/2$, where $E_F$ is the Fermi energy of the (non-exchange split) doped graphene. This means that as the Fermi energy, $E_F$, is shifted further away from the Dirac point by doping or gating, the magnitude of $P$ decreases. In our case, for a doping-induced $E_F \approx$ -150 meV for Py and $E_F \approx$ +30 meV for Co (taking into account finite 'back screening' in FM-G-G-FM sandwiches, as determined in our electrostatics modelling, Supplementary Note 6), we can assume $|P_{Co}| \approx |5P_{Py}|$. With the experimentally obtained $|MR| \approx 0.5$ % at zero bias (after annealing), we then obtain $P_{Co} \approx 10\%$ and an estimated exchange splitting $E_{ex} \approx 6$ meV. This is a similar order of magnitude as the experimentally measured value of 2 meV for graphene on EuS [8], although larger than 25 μeV found for graphene on YIG in ref. [6] (here we refer to the exchange energy g·μ$_B$×|**B**$_{exch}$| [6] resulting from the reported exchange fields $B_{exch}$ ~ 0.2T and 14T for graphene on YIG [6] and EuS [8], respectively).

In conclusion, our results reveal a new role played by graphene in MTJs. In place of the earlier discussed 'K-point spin filtering' suggested for both FM-graphene-FM and FM-hBN-FM epitaxial-quality interfaces [10-12], we find that the major role is played by a combined effect of metal-induced doping and exchange-proximity induced spin polarization in graphene, making it a tunable source of spin-polarised electrons rather than a tunneling barrier. This conclusion opens up a new way to create devices where magnetoresistance can be controlled by the size and polarity of charge transfer in proximitised graphene. More ambitiously, one may achieve electrostatic control over MR by implementing G-hBN-G in van der Waals heterostructures with ferromagnetic insulators, if a larger exchange splitting in graphene is obtained as predicted by theory [7]. At optimal electrostatically controlled doping, this can drive proximitised graphene into the half-metallic state as illustrated in Fig. 6, offering exciting prospects for graphene applications in MR devices.

## METHODS

All devices were made using a 4-step fabrication process shown schematically in Supplementary Fig. S1. To produce an aperture in the double-sided $SiN_x$(100 nm)/Si(200 µm)/$SiN_x$(100 nm) wafer used as support for FM-G(hBN)-FM' structures, we first used plasma etching (Oxford PlasmaLab System 100 ICP) with a Shipley 1813 photoresist etch mask to remove a ~300 x 300 µm area of $SiN_x$. This was followed by wet etching of the exposed Si in a KOH solution (30 wt. % in $H_2O$ at 90 °C) until a 100 nm thick $SiN_x$ window was produced, with a typical area ~25 µm x 25 µm. After that a 3.5 µm diameter circular aperture was drilled in the $SiN_x$ window using a dual-beam FEI Nova NanoLab 600 focused ion beam (FIB) with 30 kV Ga+ ions at a current of 0.1 nA (prior alignment was done with the SEM column to minimize damage). By approaching the $SiN_x$ membrane from the bottom side (Supplementary Fig. S1) FIB etching produced a concave profile of the aperture walls, ensuring that the thin metal film deposited in the next step was continuous, minimizing any complications due to e.g. fracture. In the second step, a few-layer graphene (or hBN) flake was mechanically exfoliated onto a PMMA membrane. The number of layers, N, was estimated from optical contrast and later confirmed using Raman spectroscopy. The flake was then transferred onto the aperture using a dry transfer method [23,35]. To ensure good adhesion between graphene (or hBN) and $SiN_x$, we selected flakes much larger than the aperture (minimum area of ~100 µm$^2$). In addition, the wafer was heated to 60° C during the transfer, which was followed by a 45-second annealing at 130° C. The PMMA was then detached from the flakes by gently dipping the sample in acetone (10 min) and the residual solvent was removed in isopropyl alcohol and hexane baths. The suspended flake was characterized by differential interference contrast (DIC) microscopy, atomic force microscopy (AFM) and Raman spectroscopy – information about the number of atomic layers was extracted from Raman spectra (for graphene) or AFM data (for hBN). For devices containing both hBN and graphene, we used a transfer procedure as described in ref. [23]. In the third step, Co(20 nm)/Ti(6 nm)/Au(50 nm) were deposited onto graphene(hBN)/$SiN_x$ through an Al shadow mask in an e-beam evaporator (Leybold L560 Universal Coating System) at a base vacuum of $10^{-6}$ mbar and a low deposition rate (0.03 nm/s) to reduce stresses. In the final fourth step, we evaporated $Ni_{0.8}Fe_{0.2}$(20 nm)/Ti(6 nm)/Au(120 nm) on the bottom side using the same conditions as for the top side deposition.

To image and characterize the samples at various fabrication steps we used optical, atomic force (AFM), scanning electron (SEM) and Raman microscopies. AFM imaging was done in air using a Bruker Dimension FastScan in PeakForce Tapping mode and ScanAsyst-Fluid+ tips at a 0.3 µm·s$^{-1}$ scan rate and 1 nN peak tapping force. For SEM imaging and Raman spectroscopy, a Zeiss Ultra Plus SEM and a Renishaw inVia Raman Microscope (laser excitation at 532 nm, laser spot power at sample area 0.17 mW) were used. SEM imaging was done only for test samples and never the devices, as the electron beam is known to introduce surface contamination. Device characterization is discussed in detail in Supplementary Notes 2 and 4.

**Acknowledgements.** We acknowledge support from the EC-FET Graphene Flagship, grant agreement no. 604391 and from the Marie Curie Initial Training Network "Spintronics in Graphene" (SPINOGRAPH), grant 607904.

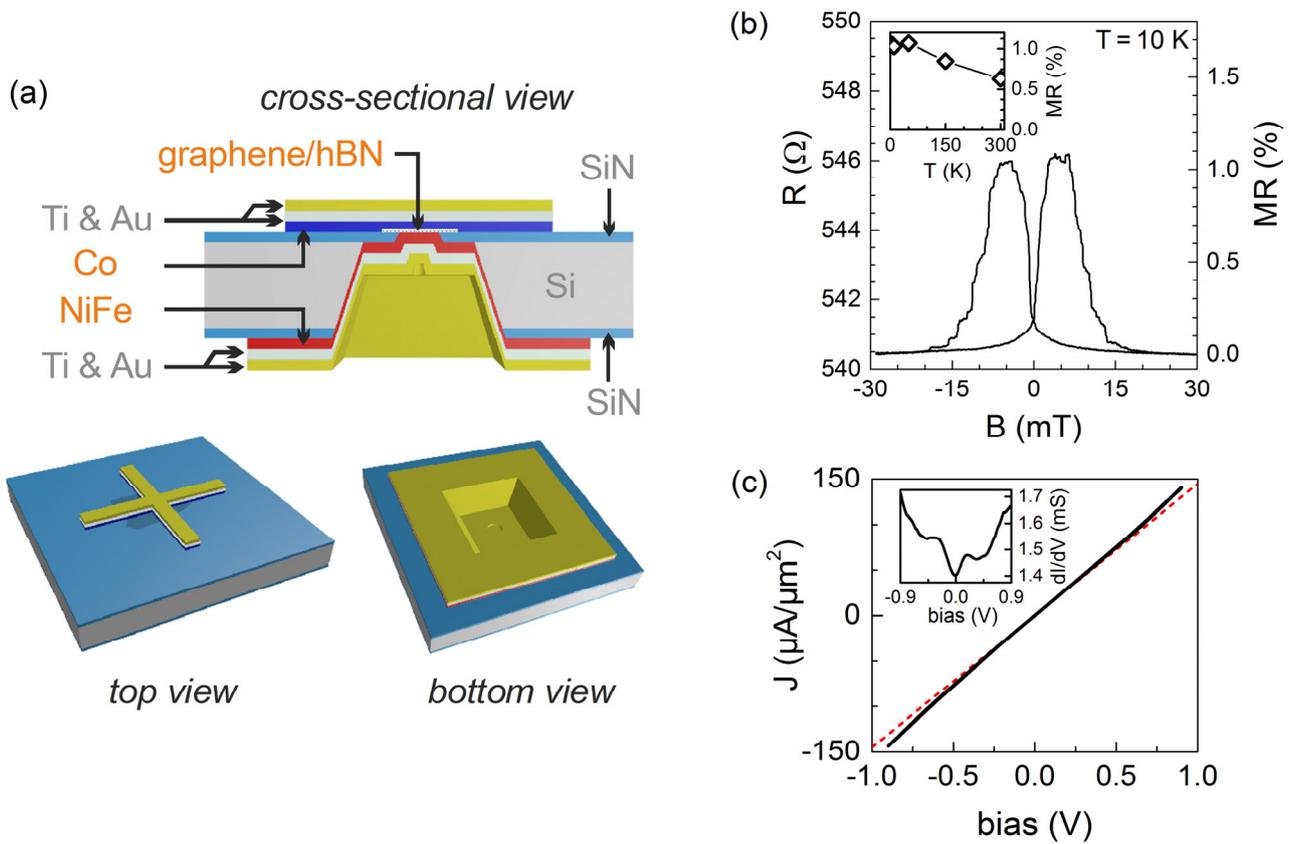

**FIG. 1. Sample design and magnetoresistance of the reference Co-hBN-Py device. (a)** Design of the samples. Ferromagnetic Co and Ni$_{0.8}$Fe$_{0.2}$ films (each 20 nm thick) were deposited on the top and bottom side, respectively, of graphene (or hBN) suspended over a 3.5 μm diameter aperture in a 100 nm thick SiN$_x$ window. On the top side, a Greek cross geometry was used. On the bottom side, a larger square area was covered by the NiFe-Ti-Au film. **(b)** Resistance $R$ of the hBN-based device vs the applied in-plane magnetic field $B$ ($T$=10 K); MR ≈ 1%. Shown behavior is typical for all $T$ up to 300 K. **Inset:** temperature dependence of the MR (◊). The line is a guide to the eye. **(c) Main panel:** Current-voltage characteristic of the hBN-based device at $T$ = 10K (black solid line). Its tangent line at zero bias (red dashed line) is shown to illustrate a slight bending of the curve for higher bias, indicative of tunneling behavior. The inset shows the corresponding d$I$/d$V$ curve.

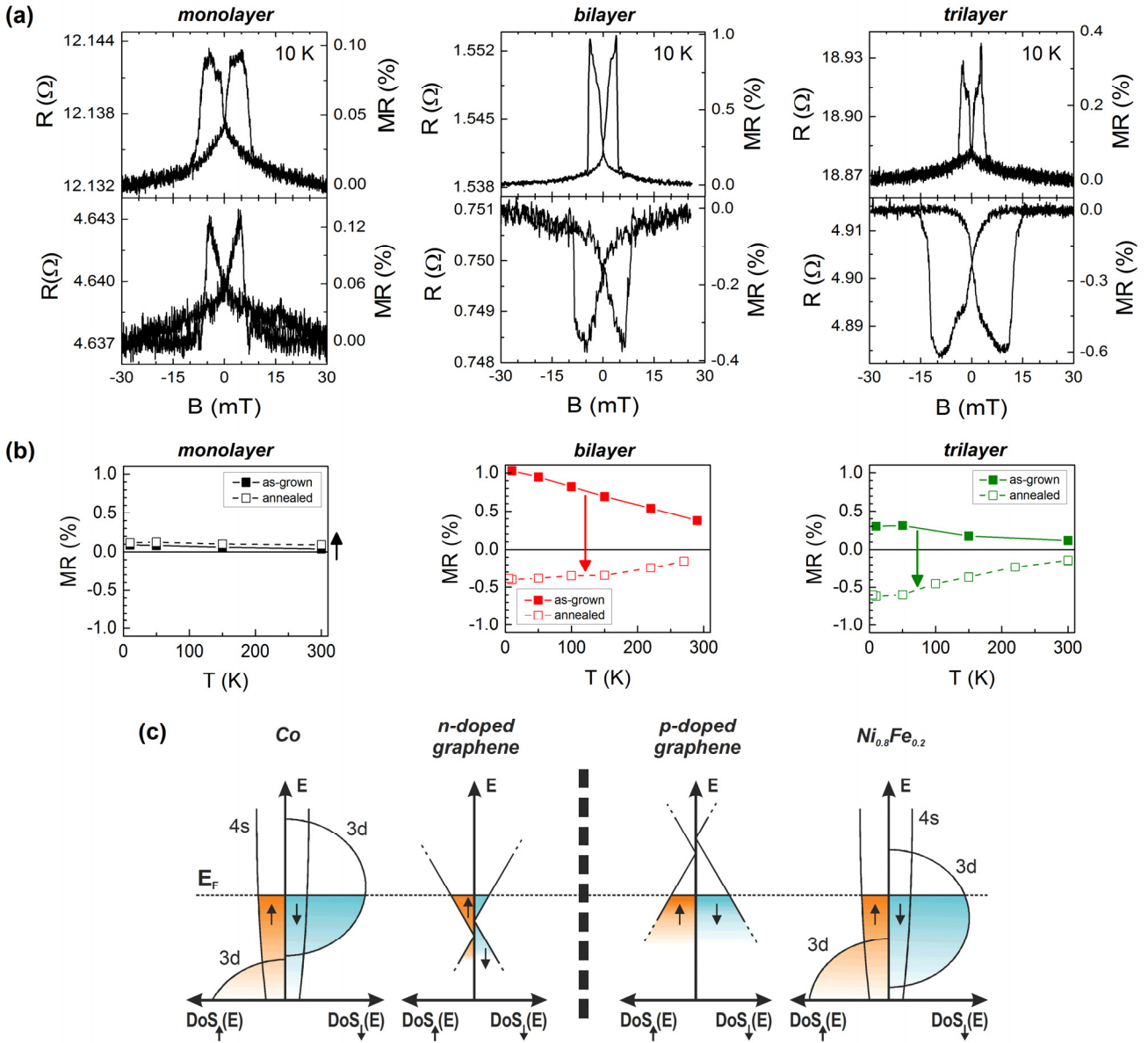

**FIG. 2. Magnetoresistance of graphene-based vertical junctions (Co-G-Py) as a function of the number of layers. (a)** Low-temperature MR traces for mono-, bi- and trilayer graphene devices before and after annealing (top and bottom panels, respectively). For $N= 2$ and 3, MR changes sign, which is accompanied by an increase in coercivity of the Co electrode. **(b)** Temperature-dependent MR of mono-, bi- and trilayer graphene devices before (□,□,□) and after (■,■,■) annealing. Arrows indicate the change in MR after annealing. **(c)** Schematic spin-dependent density of states, $DoS_{\uparrow/\downarrow}(E)$, for the constituent layers of a device with $N \geq 2$ (see text). The Fermi levels of graphene layers adjacent to the FMs are shifted due to n-doping and p-doping from Co and Py, respectively. Proximity-induced exchange splitting results in different DoSs for graphene's spin-up and spin-down carriers at $E_F$. The thick dashed line in the middle indicates decoupling of the van der Waals bonded graphene layers as they conform to the FM films (see text).

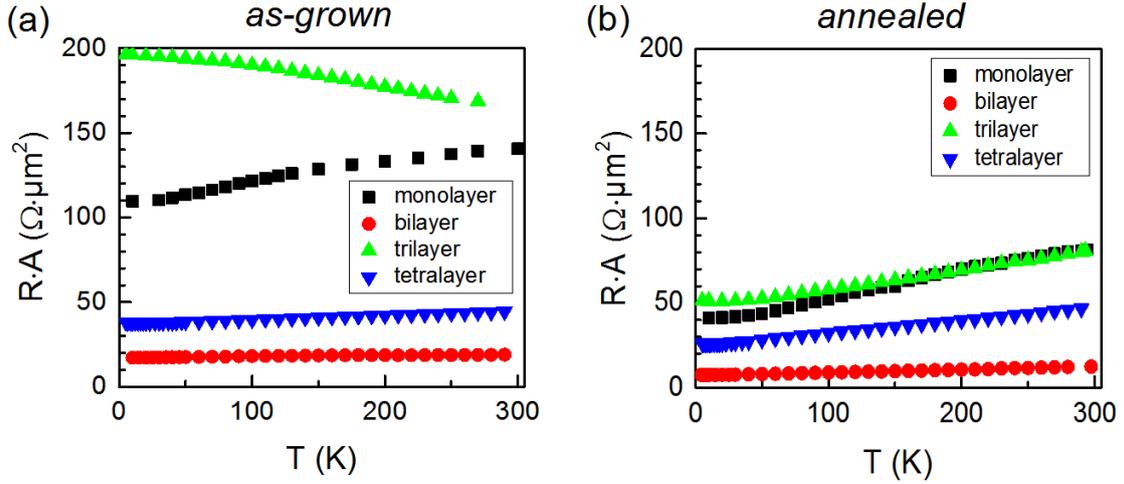

**FIG. 3. Temperature dependence of the RA product before and after annealing for Co-G-Py junctions.**
**(a)** Temperature dependences of the RA product for $N=1$ (■), $N=2$ (●), $N=3$ (▲) and $N=4$ (▼) graphene devices before annealing. No correlation can be discerned between $N$ and the device resistance and neither of the samples shows a metal-like RA($T$) dependence. **(b)** Temperature dependences of the RA product for the same devices after annealing. All devices show metal-like behavior, with a significant reduction in RA values compared to their as-grown state. The rather large statistical spread of data narrows down after annealing (we attribute the RA distribution before annealing to the presence of numerous voids at the interface, see text).

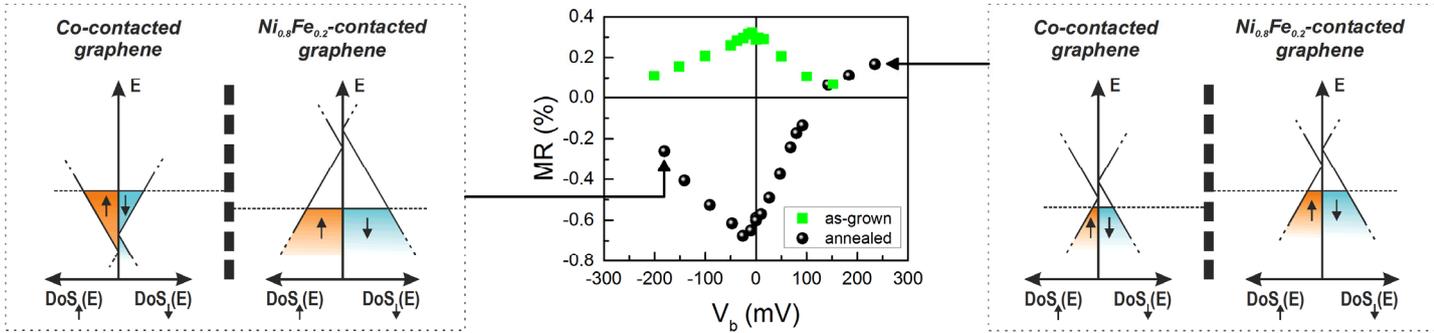

**FIG. 4: Bias dependence of the magnetoresistance before and after annealing for Co-G-Py devices.**
**Central panel:** Experimental dependence of MR on the bias voltage, $V_b$, for a trilayer graphene device before (■) and after (●) annealing. For the annealed device, MR changes sign from negative to positive at $V_b \approx +120$ mV (corresponding to carriers flowing from Py to Co). **Right panel:** Schematic DoS in the two doped and proximitised graphene 'electrodes' at maximum positive $V_b$ (see text). As $E_F$ of graphene adjacent to Co shifts into the valence band at $V_b > +120$ mV, spin-down carriers have a higher DoS in both graphene 'electrodes' and MR becomes positive (c.f. Fig. 2c). **Left panel:** Schematic DoS in the two graphene 'electrodes' at maximum negative $V_b$. No sign change of MR is expected, in agreement with experiment, as negative $V_b$ does not change the prevalent spin type at $E_F$ for the two graphene electrodes.

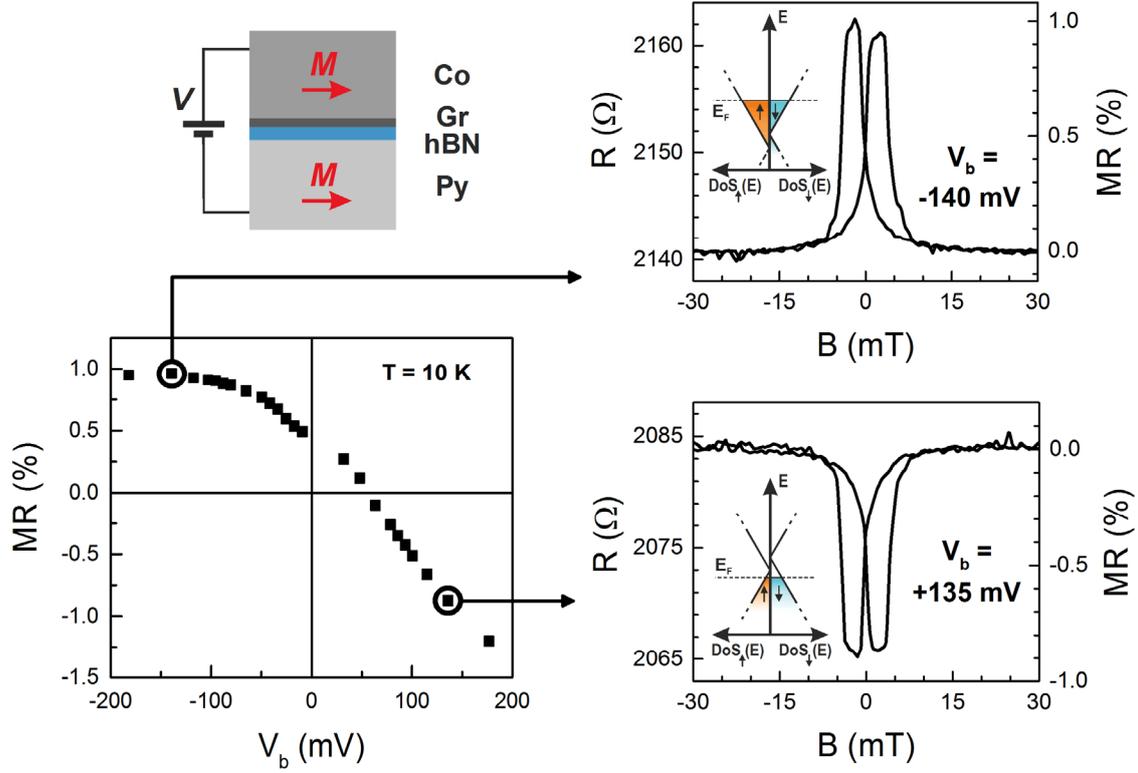

**FIG. 5: Sign reversal of MR under bias voltage for Co-G-hBN-Py devices.** Bottom-left panel: MR as a function of bias voltage, $V_b$, for a Co-G-hBN-Py device with the spacer between Co and Py made of a monolayer graphene and a bilayer hBN (sketch of the device is shown on the top-left). The panels on the right show magnetoresistance traces for maximum positive and negative $V_b$, as indicated. The MR sign reversal occurs at $V_b \approx +60$ meV, corresponding to a shift in graphene's Fermi energy from the conduction to valence band for increasing $V_b$ (Supplementary Note 6). The MR was acquired using DC electrical characterization.

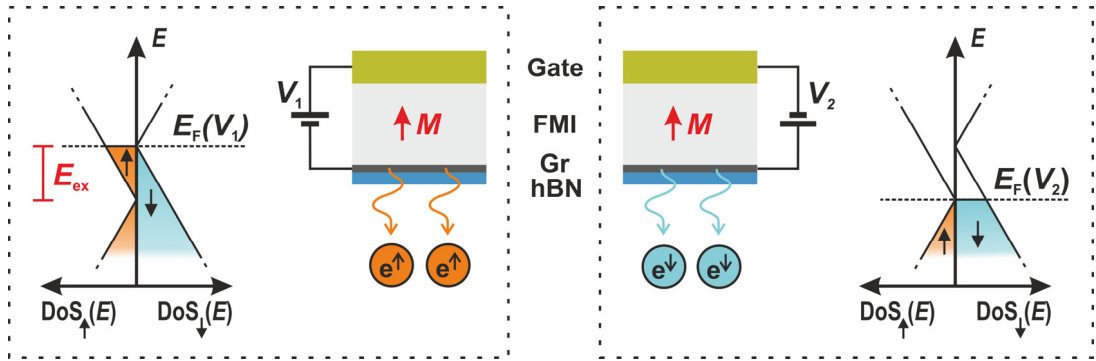

**FIG. 6: Interplay between proximity-induced exchange splitting and charge transfer in MTJs.** Shown is a sketch of the optimal conditions for creating fully spin-polarised injectors controlled by electrostatic doping of graphene in proximity to a ferromagnetic insulator (FMI). Gate voltages $V_1$ and $V_2$ provide the condition $E_{F\pm} = \pm\mu_B \times |\mathbf{B}_{exch}|$ for the half-metallic regime in graphene.

# Supplementary Information

Magnetoresistance of vertical Co-graphene-NiFe junctions controlled by charge transfer and proximity-induced spin splitting in graphene

P. U. Asshoff, J. L. Sambricio, A.P. Rooney, S. Slizovskiy, A. Mishchenko, A.M. Rakowski, E.W. Hill, A.K. Geim, S. J. Haigh, V. F. Fal'ko, I. J. Vera-Marun, and I. V. Grigorieva*

**Supplementary Note 1. Device fabrication.**

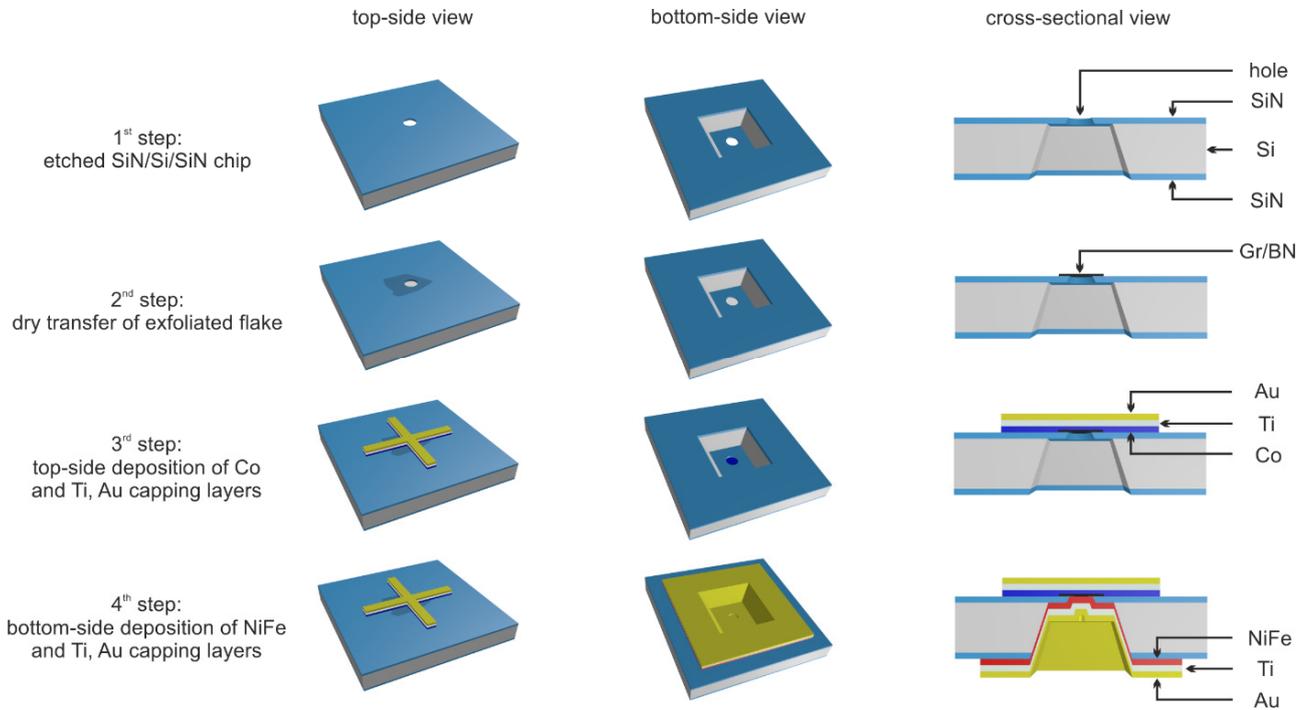

**Supplementary Figure S1. Schematics of device fabrication.** Details of each fabrication step are given in Methods.

Compared to earlier experimental studies of vertical FM-graphene-FM devices that used FM deposition directly onto graphene (either suspended [S1] or exfoliated onto a polymer membrane [S2]), we used fewer fabrication steps, avoided using more aggressive solvents, such as NMP, and used thinner ferromagnetic films, thereby reducing stresses in the devices. Furthermore, FIB etching from the back side of the device made it possible to achieve an aperture profile that ensured that both FM films (Co and Py) were continuous across the aperture. This, together with the relatively large aperture sizes ensured a clear switching behavior of the FM films, therefore avoiding complications due to e.g. magnetic domain configurations. The device geometry and dimensions were the same in all our samples and only the number of graphene layers was varied. This allowed us to unambiguously relate the measured magnetoresistance to the properties of the graphene barrier. Furthermore, the architecture of our devices allowed us to avoid the well-known issues regarding edge conduction: In all devices the edges of graphene spacers and the FM electrodes are very far away from the junction area.

**Supplementary Note 2. Device characterization.**

To characterize our samples at different stages of fabrication we used optical, atomic force (AFM), scanning electron (SEM) and Raman microscopies. To assess the roughness and quality of graphene spacers, suspended areas of the devices were imaged using AFM. This showed that the surface of the suspended graphene was clean and very smooth, without any signs of polymer residue. The RMS roughness for a typical graphene monolayer was 0.47 nm (Supplementary Fig. S2a). We note that this is an important improvement compared to the earlier report [S2] where a similar fabrication method was used but required treatment with NMP-based solvents. As a test, we have prepared several samples using a PMMA-based transfer technique followed by the removal of resist with NMP. This resulted in much dirtier surfaces, typically showing residue with oil-like adhesion and viscosity.

To verify the number of graphene layers in the transferred exfoliated graphene, we used Raman spectroscopy. Typical spectra for suspended mono-, bi-, tri- and tetralayer graphene are shown in Supplementary Fig. S2b. The high quality of our graphene membranes was confirmed by Raman maps acquired with a resolution of 0.7 μm, revealing the absence of the D peak everywhere in the suspended areas.

The AFM image of a typical graphene device after the deposition of both FM electrodes is shown in Supplementary Fig. S2d. The corresponding cross-sectional profile shows a smooth surface and a 'Mexican hat' shape typical for all our devices. This shape results from bulging in the graphene area under the weight of the relatively thick (146 nm) metal film and bending of the SiN window around the aperture under the weight of a thinner (76 nm) metal film. The AFM image of the finished bilayer hBN device in Supplementary Fig. S2e shows a rougher surface, probably due to island formation during Co growth on hBN as reported earlier [S3]. Importantly, the more irregular growth of Co on hBN did not affect the performance of our devices as both the Co layer and the capping Ti/Au layers are much thicker than the average roughness.

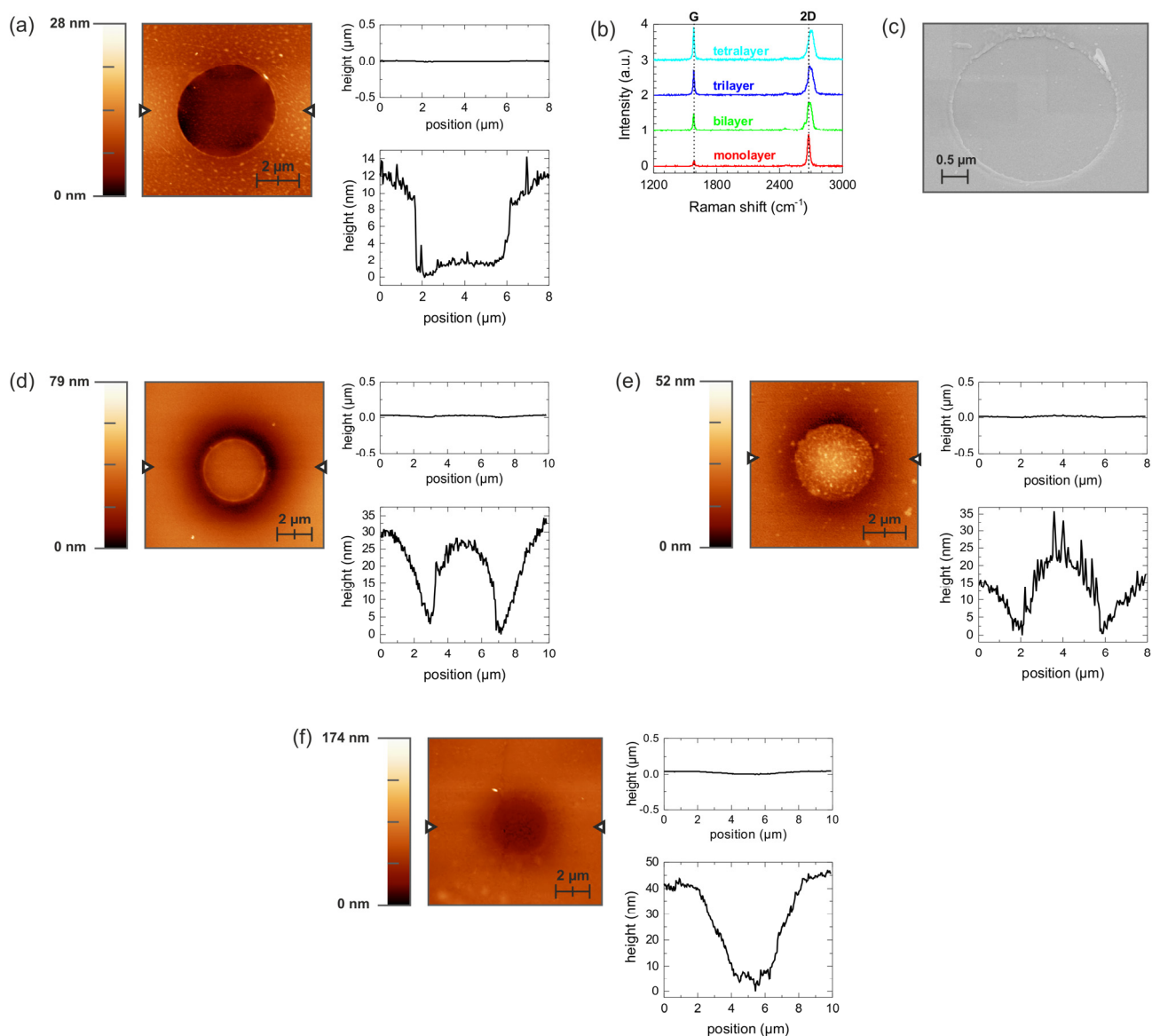

**Supplementary Figure S2. Characterization of our devices at different steps of fabrication. (a)** AFM image of a suspended monolayer graphene flake before the deposition of metal electrodes (step 2 in Supplementary Fig. S1). The size of the flake is much larger than the aperture area. A height profile corresponding to the cross section indicated by the triangles is shown on the right in two graphs showing the same data but on different scales – while the upper graph uses a uniform scale, showing an essentially flat height profile, the lower graph reveals nanoscale features. Graphene sags and partially adheres to the sides of the aperture as is common for suspended atomically thin crystals. **(b)** Raman spectra of suspended mono-, bi-, tri- and tetralayer graphene. The D peak is absent indicating high quality of graphene spacers in all our devices. **(c)** Scanning electron microscope (SEM) image of a finished bilayer hBN device taken from the bottom side after the deposition of both FM films (step 4 in Supplementary Fig. S1). The bottom FM film (Py) is continuous across the step between the ring-shaped indentation (corresponding to the thickness of the supporting $SiN_x$ membrane) and outer areas, ensuring good electrical contact. The image was taken under 30° tilt. **(d)** AFM image of a suspended bilayer graphene after metal deposition on both sides (step 4 in Supplementary Fig. S1) and the height profiles as in (a). Sagging (bending) of the supporting $SiN_x$ membrane under the weight of the top FM film (Co) and of graphene under the weight of the thicker bottom FM (Py) are clearly seen. **(e)** AFM image and height profile as in (d), but for an hBN device. **(f)** Typical AFM image of an annealed graphene

device and the corresponding height profile. Relaxation of strains is evident as the 'Mexican hat' profile disappeared and instead a more natural sagging of the suspended area occurs. All devices shown have a junction area of ≈11 µm$^2$.

To characterize the effect of annealing on the microstructure of the FM layers at the FM/graphene interface, we fabricated test devices with monolayer graphene where only one FM layer (either Co or Py) was deposited on the top side of the suspended graphene (Supplementary Fig. S3). All process parameters used for FM deposition in these test devices and subsequent annealing were exactly the same as for the devices used to study the magnetoresistance, to ensure comparable characteristics of the FM films. As graphene is effectively transparent to the electron beam, it was possible to see the graphene/FM interface, including the shapes of individual grains in our polycrystalline films and the boundaries between them. Comparison of the SEM images before and after annealing revealed clear signs of rearrangements of the crystallites after annealing. For example, some of the 'cracks' between NiFe grains that were visible on the as-grown samples became wider after annealing, indicating relaxation of strains as discussed in the main text (Supplementary Fig. S3c,d). For Co films, fewer and less defined 'cracks' are seen after annealing, indicating a smoother, less corrugated surface. At the same time, there was no increase in the size of the crystallites, as could be expected if re-crystallization had taken place. We note that SEM imaging was only used for the above test devices and never for the devices used for transport measurements, as SEM imaging is known to induce surface contamination.

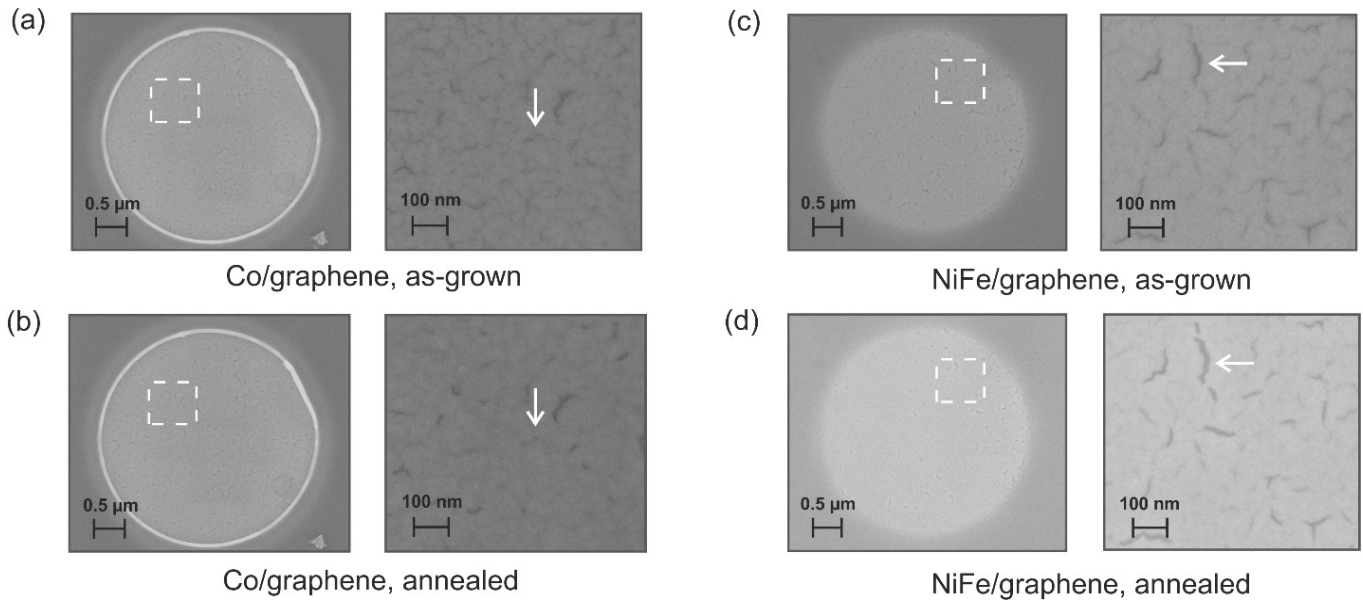

**Supplementary Figure S3. SEM imaging of Co and NiFe films through graphene, before and after annealing.** All images were taken from the bottom side of the device. Graphene is effectively transparent to the electron beam, which allows a detailed view of the structure of the ferromagnetic films. **(a,c)** SEM images of Co and NiFe films, respectively, taken before annealing (as-grown). The thickness of both Co or NiFe was 20 nm, followed by capping layers of Ti(6 nm)/Au(50 nm). **(b,d)** SEM images of the same Co and NiFe films as in (a,c) after annealing in Ar/H$_2$ atmosphere at 300°C for 3h. The areas marked by dashed lines on the left images of each panel are shown at higher magnification on the right. Arrows indicate areas where changes due to annealing are clearly seen, such as 'cracks' between individual grains in our polycrystalline films.

To investigate the mechanical strength of the bonding between graphene and the FM films, we prepared several samples consisting of graphene mechanically exfoliated onto Si/SiO$_x$ substrate and covered by thin Co/Ti/Au or NiFe/Ti/Au films. The films were of the same thicknesses and evaporated under the same conditions as those used for our TMR devices. Then the FM films were peeled off the substrate using a strongly adhesive tape to see which interfaces have the weakest adhesion. In all cases, the Au/Ti/FM/graphene stack completely came off of the substrate, leaving the bare SiO$_x$ surface. A similar behavior was found for few-layer graphene/FM stacks. This demonstrated a very strong adhesion between graphene and the FM films, stronger even than the adhesion between graphene and SiO$_x$ [S4], supporting our expectation that graphene layers adjacent to the polycrystalline FM films conform to the corrugated surfaces of the latter.

**Supplementary Note 3. Magnetotransport measurements and *I-V* characteristics.**

All FM-graphene-FM' devices exhibited linear current-voltage characteristics over a large voltage range, irrespective of the number of layers (Supplementary Fig. S4). *I-V* characteristics were linear both before and after annealing.

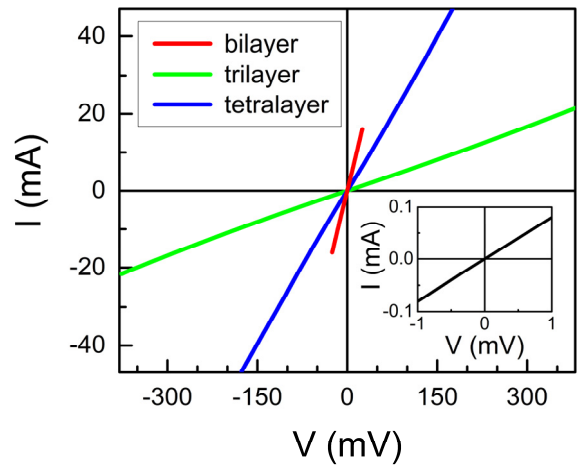

**Supplementary Figure S4.** *I-V* characteristics of FM-graphene-FM' devices. Shown are typical *I-V* characteristics of 'as-grown' devices (before annealing) with bi- (—), tri- (—) and tetralayer (—) graphene separating the FM films. The inset shows *I-V* characteristic for a monolayer device. All data acquired at 10 K.

As described in the main text, all our devices with N≥2 showed a consistent behavior after annealing, i.e., a decrease in junction resistance. Devices with N= 2 and 3 also showed a change of sign of MR (from positive to negative) but for N=4 MR decreased to almost zero but its sign reversal was not observed (Supplementary Fig. S6). There are at least two likely reasons for the observed difference between devices with N = 2,3 and N = 4: (i) Less effective conformation at the interfaces between the FM films and the tetralayer graphene: The latter is less flexible and cannot conform to the corrugated FM film as well as the thinner bi- or trilayer. Indeed, the changes in the junction resistance for N=4 were much smaller than for N=1-3 – this is seen clearly in Fig. 3 in the main text. (ii) In the likely case that N=4 layer graphene decouples into two bilayers (BLG), metal-induced doping of BLG leads to smaller shifts of the Fermi level, $E_F$, compared to the monolayer. This is due to the difference in the density of states near the neutrality point between bilayer and monolayer graphene. As a result, DoS in both bilayers is likely to be low and opposite doping polarities, as in metal-doped graphene monolayers (n- and p-doping for the Co- and Py-side, respectively) are unlikely to be realized. This would result in smaller spin polarizations in graphene bilayers adjacent to the FMs, very small MR and the absence of (or an incomplete) MR sign reversal, in agreement with our observations.

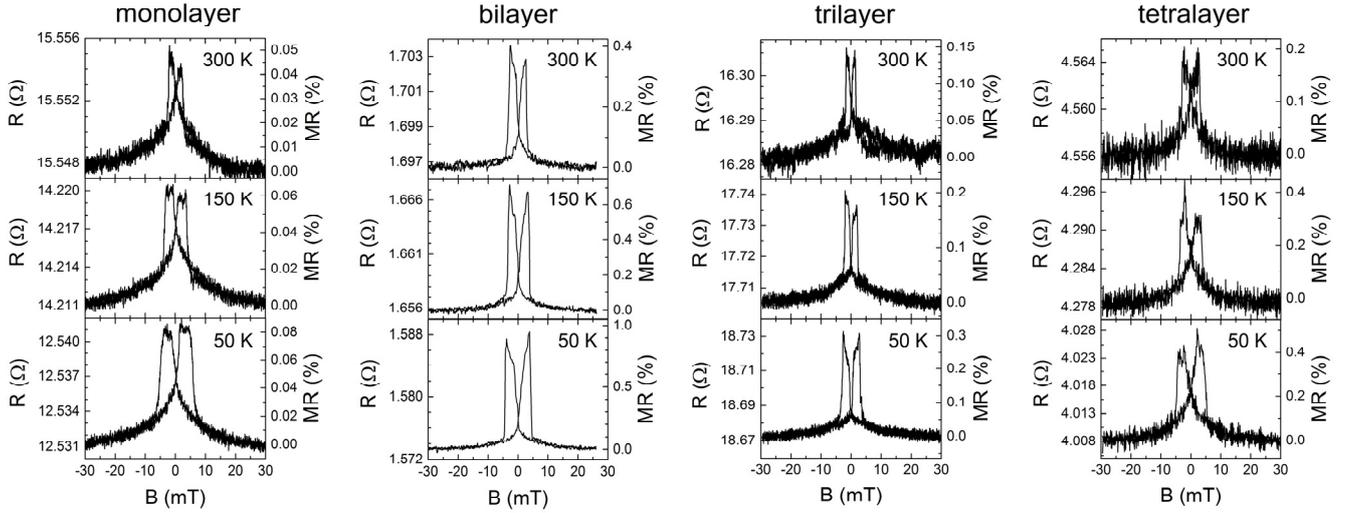

**Supplementary Figure S5. Magnetoresistance of as-grown graphene devices at higher temperatures.** Shown is the magnetic field dependent resistance, $R(B)$, for mono-, bi-, tri- and tetralayer graphene devices. The behavior is typical for vertical TMR devices (see main text). The coercivity of the Co layer increases monotonically with decreasing temperature leading to more clearly separated switching of Co and NiFe and broader MR features.

It was important to confirm that the anisotropic magnetoresistance (AMR) effect (the dependence of a ferromagnet's resistivity on the angle between the applied current and the magnetization) was negligible in our measurements. To this end, we used the device geometry (cross-shape of the top Co film) and ensured that the applied magnetic field was parallel to one arm of the cross; the current was then applied along this arm. As a control, we also applied the current through the perpendicular arm of the cross. In both cases the MR features were identical in magnitude and shape. We performed a similar procedure with bottom contacts resulting in currents flowing at different angles with respect to the magnetic field and did not see a difference in MR.

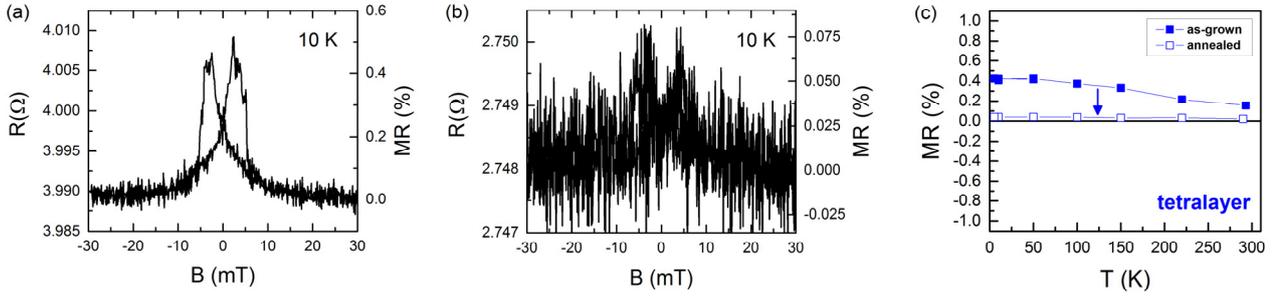

**Supplementary Figure S6: Magnetoresistance of a tetralayer graphene device. (a,b)** Low-temperature magnetoresistance sweeps before and after annealing, respectively. **(c)** Temperature-dependent MR before (■) and after (□) annealing; the arrow indicates the MR change due to annealing.

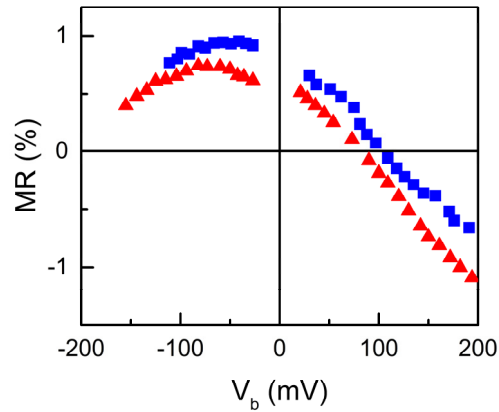

**Supplementary Figure S7: Bias dependence of Py-hBN-Gr-Co devices.** Low-temperature magnetoresistance as a function of bias voltage $V_b$ for two devices different from the one presented in Fig. 5 of the main manuscript. Triangles (▲) are data for a Py-hBN-Gr-Co device with 2 hBN layers (10 K), squares (■) for a device with 3 hBN layers (20 K). A sign reversal of MR occurs for $V_b \approx +80$ mV and $+105$ mV, respectively. The MR was acquired using DC electrical characterization.

**Supplementary Note 4. Cross-sectional STEM analysis.**

To characterize the interfaces between few-layer graphene and the ferromagnetic films and to assess the effect of annealing, we used cross-sectional analysis in a scanning transmission electron microscope (STEM). Unannealed and annealed devices were prepared for STEM cross–sectional imaging analysis in a Helios 660 focussed ion beam (FIB) microscope using the well-known *in situ* 'lift-out' method [S5] and low kV ion beam polishing [S6,S7]. With this approach a thin (~65 nm) lamella window of material is removed from the active area of the Co- G-Py junction (perpendicular to the graphene basal planes and the interfaces between graphene and the ferromagnetic films). This thin slice of material is transferred to a specialist Omniprobe TEM support grid, mounted with the incident electron beam perpendicular to the plane of the lamella. The resulting STEM image shows a cross section of the active region of the device. STEM analysis was carried out using a probe side aberration-corrected FEI Titan G2 operating at 200 kV with a probe convergence angle of 21 mrad, a HAADF inner angle of 62 mrad and a probe current of ~90 pA. Energy dispersive x-ray (EDX) spectrum imaging was performed in the Titan using a Super-X four silicon drift EDX detector system with a total collection solid angle of 0.7 srad. The multilayer structures were oriented along graphene's <hkl0> crystallographic direction by taking advantage of the Kikuchi bands of the crystalline metal layers.

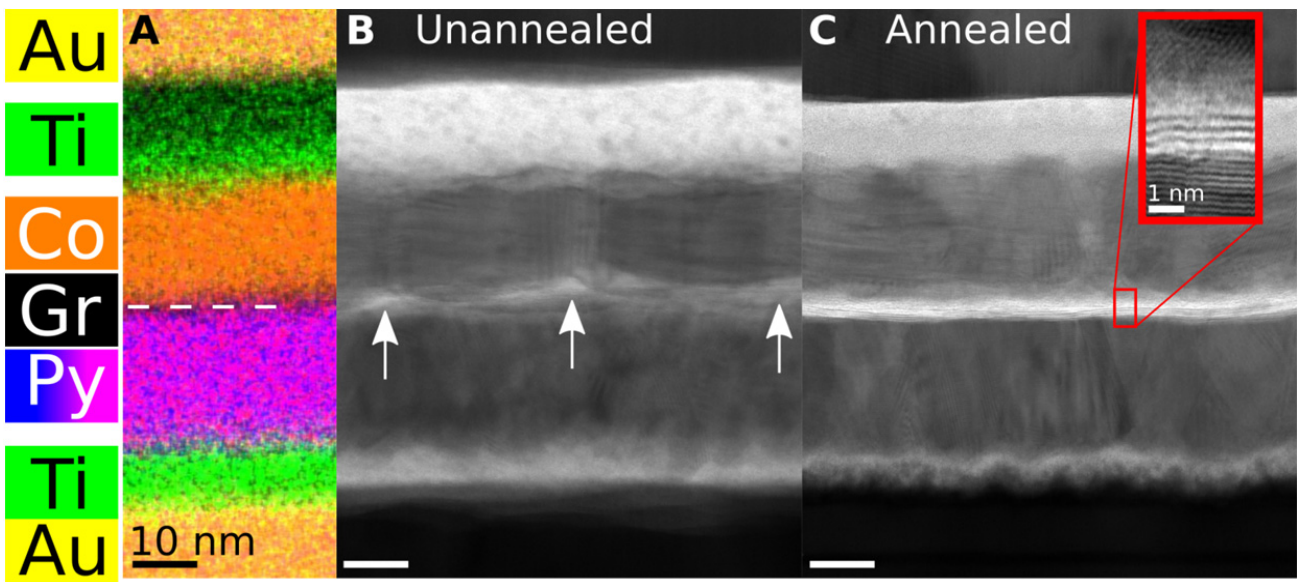

**Supplementary Figure S8: STEM analysis of device cross sections**. **(A):** An energy dispersive x-ray spectroscopy (EDXS) map showing the elemental distribution in an active region of the device. **(B)** STEM bright-field image of a typical area in the active region of an unannealed device. Small but significant voids are clearly seen around graphene, some examples are indicated by arrows. The voids arise due to the rough surfaces of both Py and Co. **(C)** STEM bright-field image of an equivalent area in an annealed device. Here the surfaces of both Py and Co have become uniform and much flatter, resulting in excellent contact with trilayer graphene, as shown by the high magnification image in the inset.

Energy dispersive x-ray spectroscopy (EDXS) data in Supplementary Fig. S8A,D shows the typical elemental distribution of the devices, which was found to be unchanged after annealing. It is clear that all components have well defined (on the scale of EDXS spatial resolution), clean interfaces resistant to interlayer diffusion. Atomic-scale STEM imaging (Supplementary Fig. 8B,C) revealed a profound change in the quality of the interfaces between graphene and the ferromagnetic films: While significant voids, up to several nm, are present between graphene and the FM films (either on Co or Py side) in as-prepared devices, annealing causes the voids to disappear and the surfaces of both Py and Co to become flatter and more uniform. This simultaneously reduces the tortuosity of graphene and allows it to adhere closer to Co and Py. To quantify the minimum distances between the graphene planes directly interfacing Co and Py crystallites, measurements between relevant atomic fringes were acquired from high resolution bright field STEM images. The bright field contrast transfer was inferred by direct comparison to HAADF images of the same area and in all cases dark regions were found to correspond to the location of the graphene planes. Measurements of the spacings between the ferromagnetic films and the adjacent graphene layers in the annealed device were made across 25 different areas (for the unannealed device such measurements are meaningless, due to the presence of numerous voids). These revealed the average distance between Co (Py) and the nearest graphene plane of 0.39 nm ± 0.06 nm and 0.34 nm ± 0.09 nm, respectively. The images were calibrated to the crystalline platinum protective layer introduced during FIB sample preparation. The internal spacing between graphene layers in the few layer material was found to be 0.33 nm ± 0.06 nm, in excellent agreement with the well-known layer separation in multilayer graphene.

**Supplementary Note 5. Metal-induced doping of graphene.**

To determine the level of doping of graphene from the adjacent FM films in our devices, we prepared specially designed test samples and measured gate-dependent charge transport in the vicinity of Co and Py contacts, a technique similar to that used in ref. [S8]. To this end, a rectangular graphene flake was mechanically exfoliated onto a Si/SiO$_x$(290 nm) substrate. Co and NiFe bars were then evaporated on top of graphene at different spacings, $s$ = 0.5 μm to 4 μm, as shown in Supplementary Fig. S9d. To ensure compatibility with magnetotransport measurements in our FM-graphene-FM' devices, we used identical deposition conditions for the FM films. For each Co or NiFe contact pair separated by a spacing $s$ we measured the resistance, $R$, of the graphene channel as a function of gate voltage, $V_g$. For small $s$, doping from the FM contacts is expected to dominate the measured $R(V_g)$, whereas for larger $s$ the resistance will be largely determined by environmental doping (substrate and adsorbate induced). We therefore expect two Dirac points to emerge in $R(V_g)$, one originating from the metal-induced doping that is at a maximum directly under the contacts but also extends to a ~100-300 nm distance into the graphene channel (the exact distance depending on the metal) [S9-S12], and another one from the central area of the graphene channel. Accordingly, we designate the gate-dependent resistance of the contact-doped part of the channel and of its central region as $R_{metal}(V_g)$ and $R_{envr}(V_g)$, respectively. From the Dirac point associated with $R_{metal}(V_g)$, it is then possible to extract the shift of the Fermi level, $E_F$, due to doping from the ferromagnetic films.

Transport measurements were done in a quasi-4-terminal geometry as shown in Supplementary Fig. S9d. In the case of Co contacts, for the largest spacing $s$ = 4 μm we found a single peak in $R(V_g)$ with the Dirac point at $V_g$=+11 V (Supplementary Fig. S9a), i.e., away from the metal graphene is p-doped as is usual for environmental doping of graphene on SiO$_x$. For $s$ = 3 µm and below, a second peak developed at negative $V_g$, which became most apparent for $s$ = 0.5 µm and 1 µm. This second Dirac point, corresponding to electron doping, is due to a large part of the graphene channel being doped by the Co contacts and its resistance becoming an increasingly large part of the total $R$. To extract the carrier density associated with doping by Co contacts, we used a two-peak fitting procedure based on a phenomenological model proposed in ref. [S13]. In this model, the total carrier density for each peak is given by

$$n_{tot} = [n_0^2 + n(V_g)^2]^{1/2}, \quad (1)$$

where $n_0$ is the intrinsic carrier density and $n$ the carrier density induced by the gate voltage $V_g$ according to $n(V_g) = (C/e)(V_g - V_0)$, where $V_0$ is the gate voltage corresponding to the Dirac point and $C$ the capacitance of the SiO$_x$ dielectric layer. The total resistance $R_{tot}(V_g) = R_C + R_{\text{metal}}(V_g) + R_{\text{envr}}(V_g)$ then results from the sum of the constant contact resistance $R_C$ and the resistance of the two differently doped graphene areas, $R_{\text{metal}}(V_g) \propto (n_{\text{metal}} e \mu)^{-1}$ and $R_{\text{envr}}(V_g) \propto (n_{\text{envr}} e \mu)^{-1}$, where $n_{\text{metal}}$ and $n_{\text{envr}}$ are determined by eq. (1). The positions, $V_0$, of the two Dirac points extracted from the fits for different $s$ are shown in Supplementary Fig. S9c. The fact that both Dirac points remain at approximately the same respective $V_g$ for all $s$ confirms the validity of our approach.

To relate $n_{\text{metal}}$ and the associated shift in $E_F$ due to doping we can use [S14]

$$n(E_F) = \frac{E_F^2}{\pi(\hbar v_F)^2}. \quad (2)$$

On the other hand, the gate-induced carrier density, $n(V_g)$, can be found as

$$n(V_g) = V_g \varepsilon_0 \varepsilon_r / (d \cdot e), \quad (3)$$

where $v_F \sim 10^8$ cm/s is graphene's Fermi velocity, e the electron charge, $\varepsilon_0$ the permittivity of the free space and $\varepsilon_r = 3.9$ and $d$ = 290 nm the permittivity and thickness of the SiO$_x$ layer, respectively. Using (2), (3) and $V_0 \approx$ -11 V for the Dirac point associated with doping from the Co contacts, we find doping-induced $E_F \approx$ +105 meV, i.e. graphene is n-doped.

A similar analysis for the sample with NiFe contacts showed that graphene in contact with NiFe is p-doped, with a larger Fermi level shift compared to Co, $E_F \approx$ -190 meV. In this case, a single $R(V_g)$ peak at $V_0 \approx$ +10 V is found for $s$ = 3 µm and a second peak at $V_0 \approx$ +36 V develops at smaller $s$, corresponding to $E_F \approx$ -190 meV (Supplementary Figure S9b).

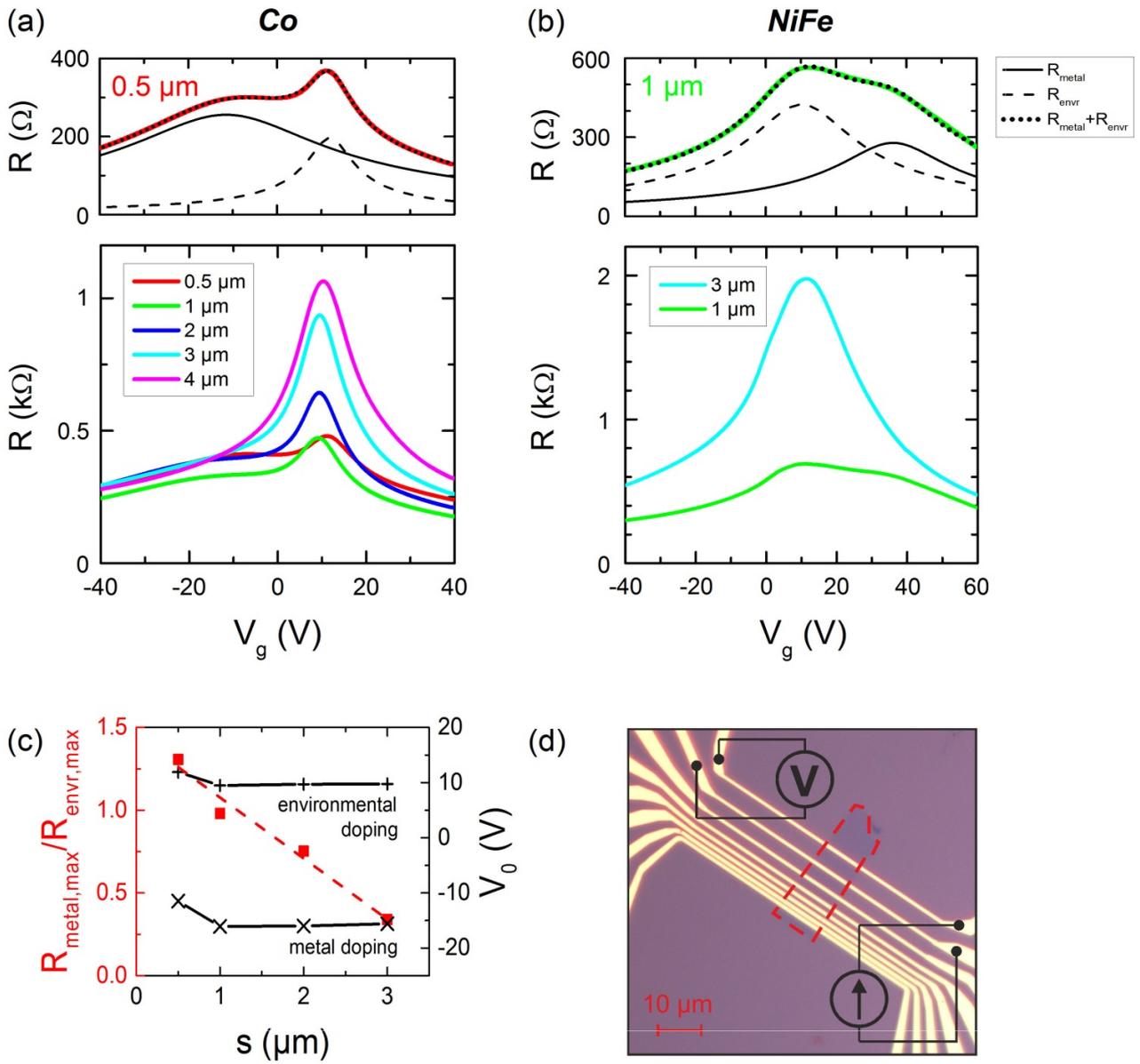

**Supplementary Figure S9: Doping of graphene by Co and NiFe. (a, b)** Resistance $R$ as a function of gate voltage, $V_g$, for graphene devices with Co and NiFe contacts, respectively. Bottom panels show measured $R(V_g)$ for different contact separations, $s$. Top panels show the results of the two-peak fitting procedure described in the text for the smallest contact separations, $s = 0.5$ μm and $s = 1$ μm for Co and NiFe, respectively. Individual peaks for $R_{metal}(V_g)$ and $R_{envr}(V_g)$ are shown by solid and dashed lines, respectively. The dotted lines show the sum of the two peaks, in agreement with the measured $R(V_g)$ shown by the solid red line. The constant contact resistance, $R_C$, obtained from the fits has been subtracted for clarity. **(c)** Positions, $V_0$, of the two Dirac points for the device with Co contacts as a function of the contact spacing, $s$ (crosses). Filled red symbols show the corresponding ratios of the maximum resistances of the two peaks. As expected, the resistance corresponding to metal doping, $R_{metal,max}$, becomes dominant at the smallest $s$. **(d)** Optical image of the device with Co contacts and the measurement geometry. Graphene is outlined by the dashed red line.

**Supplementary Note 6. Charge transfer in metal-graphene multilayers.**

To determine the doping level for graphene in Co-Gr-Gr-Py devices studied in this work, we apply a standard charge transfer model [S15] to the multilayer system sketched in Fig. 2c of the main text and in Fig. S10d. The input parameters in the model are the effective distances ($d_{Co}$ from Co to the top graphene flake, labelled "1", $d_{Py}$ from permalloy ($Ni_{0.8}Fe_{0.2}$) to the bottom graphene flake, labelled "2") and the distance $d \approx 0.4$ nm between the two graphene layers), and the work functions of monolayer graphene ($W_{Gr} \approx 4.6$ eV), polycrystalline Co ($W_{Co} = 4.4$ eV) and permalloy ($W_{Py} \approx 0.8 W_{Ni} + 0.2 W_{Fe} \approx 5$ eV) taken from refs. [S16-S21]. The values of work functions determine the difference $W_G - W_{Co} = \mu_{Co} = 0.2$ eV and $W_{Gr} - W_{Py} = \mu_{Py} = -0.4$ eV. The distances $d_{Py}$ and $d_{Co}$ are assumed to include the effect of electric field penetration into the metals, so that they can be a bit longer than the distances $d_{Co} = 0.39$ nm and $d_{Py} = 0.34$ nm between atomic layers measured using STEM (Supplementary Note 4). Below, we perform electrostatic analysis for the following choices of parameters of metal-graphene multilayers:

(a) $d_{Co} = 0.39$ nm, $d_{Py} = 0.34$ nm, $d = 0.4$ nm corresponding to experimentally measured distances between graphene and Co / Py respectively;

(b) $d_{Co} = 0.45$ nm, , $d_{Py} = 0.4$ nm, $d = 0.4$ nm taking into account the likely effect of finite electric field penetration depth into the metals;

(c) $d_{Co} = 0.39$ nm, $d_{Py} = 0.34$ nm, $d = 0.45$ nm allowing for a larger separation of graphene layers due to decoupling.

Note that for the quoted-above work function differences we find that in monolayer graphene covered by Co, n-doping corresponds to $E_F = 117$ meV for $d_{Co} = 0.39$ nm and $E_F = 122$ meV for $d_{Co} = 0.45$ nm. On the other hand, covering graphene with permalloy makes graphene p-doped, with $E_F = -207$ meV for $d_{Py} = 0.34$ nm and $E_F = -196$ meV for $d_{Py} = 0.4$ nm. These values are in good agreement with the experimentally measured graphene doping by Co and Py, as described in Supplementary Note 5. To find the Fermi energies, $E_{Fi}$ (i = 1,2), in each of the two graphene layers, we use their relation to the charge densities $\rho_i$ transferred from the metals to the naturally undoped monolayer graphene,

$$\rho_i = -eE_{Fi} \frac{|E_{Fi}|}{\pi(hv)^2},$$

These densities enter the charge balance equation,

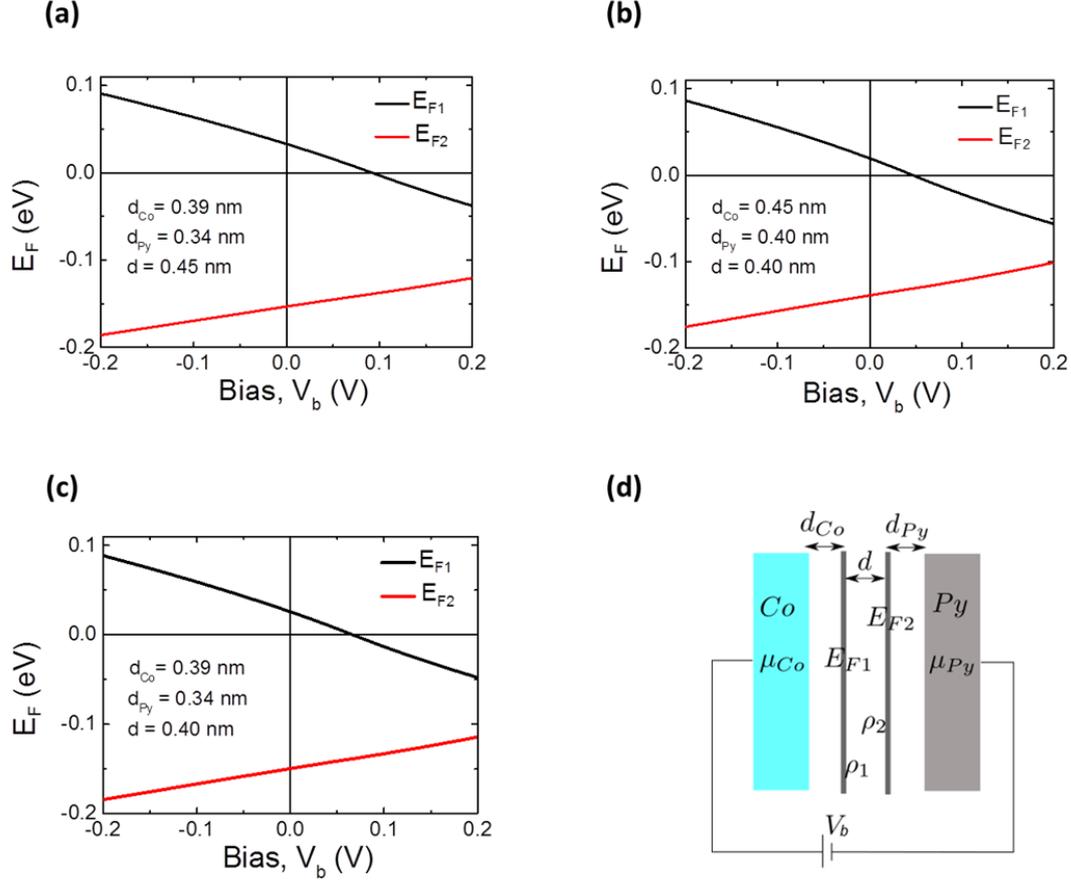

**Supplementary Figure S10: Fermi levels, $E_{F1}$ and $E_{F2}$, of graphene layers in a Co-Gr1-Gr2-Py vertical junction. (a)** $d_{Co}$ = 0.39 nm, $d_{Py}$ = 0.34 nm, $d$ = 0.4 nm corresponding to experimentally measured distances between graphene and Co/Py, respectively; the distance between the two graphene layers is taken slightly larger than graphite interlayer distance, taking into account the decoupling of graphene layers. **(b)** $d_{Co}$ = 0.45 nm, $d_{Py}$ = 0.4 nm, $d$ = 0.4 nm, taking into account the likely effect of finite electric field penetration depth into the metals. **(c)** $d_{Co}$ = 0.39 nm, $d_{Py}$ = 0.34 nm, $d$ = 0.45 nm, assuming a larger distance between the graphene layers. **(d)** Sketch of the Co-Gr-Gr-Py junction indicating all relevant distances.

$$\rho_1 + \rho_2 + \rho_{Co} + \rho_{Py} = 0,$$

where $\rho_{Co}$ and $\rho_{Py}$ are the surface charge densities on Co and Py respectively. The conditions of equilibrium between each graphene flake and the closest metal are:

$$E_{F1} = \mu_{Co} - e\rho_{Co}d_{Co}/\epsilon_0,$$

$$E_{F2} = \mu_{Py} - e\rho_{Py}d_{Py}/\epsilon_0,$$

where $\epsilon_0$ is dielectric constant of vacuum. Finally, the bias voltage applied to the whole structure is related to these parameters as

$$eV_b = E_{F2} - E_{F1} + e(\rho_{Co} + \rho_1)d/\epsilon_0,$$

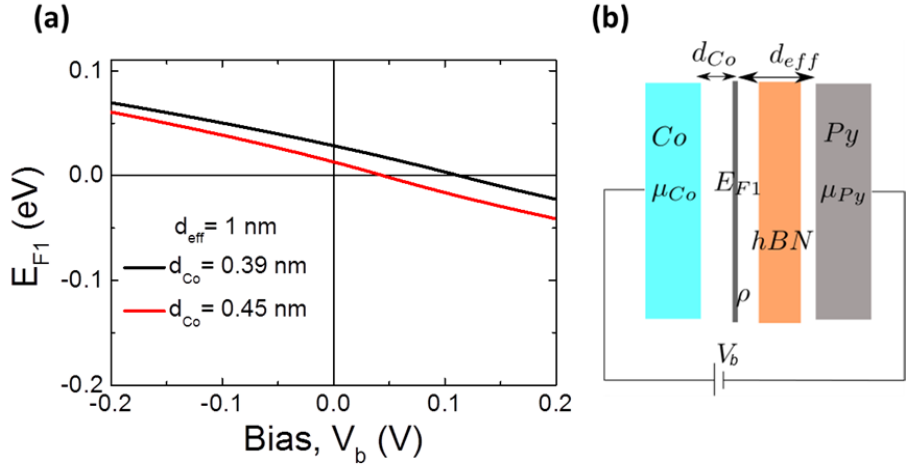

**Supplementary Figure S11. Fermi level, $E_F$, for Co-Gr-hBN-Py vertical junction.** (a) $d_{eff}$ = 1 nm, $d_{Co}$ = 0.39 nm (black line), $d_{Co}$ = 0.45 nm (dashed red) (b) Sketch of the junction.

The calculated dependences of the Fermi energies $E_{F1,2}$ in each of the two graphene layers in the TMR structure are shown in Fig. S10. These results show that for all choices of distances $d_{Co}$ and $d_{Py}$ the graphene layer adjacent to Co switches from n-type to p-type doping at $V_b \sim$ +100 mV. The bias voltage, corresponding to the change of sign of doping, increases with the distance between the two graphene layers and slightly decreases if the effective distances to the metals are increased.

Similarly, the charge transfer problem can be solved for the case of Co-Gr-hBN-Py junction (Fig. 5 of the main text). In this case, we have to take into account the dielectric constant of hBN, therefore, we introduce an effective distance,

$$d_{eff} = \frac{d_{hBN}}{\epsilon_{hBN}} + d_{Py} + d_{Gr-hBN}$$

with $\epsilon_{hBN} \approx 3.5$. For a 3-layer hBN, $d_{hBN} \approx$ 1nm, resulting in $d_{eff} \approx$ 1nm. Then the charge transfer equations take the form

$$\rho_1 + \rho_{Co} + \rho_{Py} = 0,$$
$$\rho_1 = -eE_{F1}\frac{|E_{F1}|}{\pi(hv)^2}$$
$$E_{F1} = \mu_{Co} - \frac{e\rho_{Co}d_{Co}}{\epsilon_0},$$
$$eV_b = -E_{F1} + \mu_{Py} + e\rho_{Py}d_{eff}/\epsilon_0,$$

Solving the above equations results in the bias-voltage-dependence illustrated in Fig. S11 for the same choice of $d_{Co}$ as in the analysis of Co-Gr-Gr-Py multilayers. The calculated $E_{F1}(V_b)$ displays a change of doping polarity in graphene at bias voltages $V_b \sim$ +50 to +100mV, in excellent agreement with experimental observations. The bias voltage corresponding to the change of doping polarity decreases for a larger $d_{Co}$.